\documentclass[sigconf,nonacm]{acmart}
\AtBeginDocument{}

\usepackage{nicefrac}
\usepackage{xspace}
\usepackage{colortbl}
\usepackage{amsmath,amsfonts}
\usepackage{algorithm}
\usepackage{algpseudocode}
\usepackage{graphicx}
\usepackage{textcomp}
\usepackage{xcolor}
\usepackage{tikz}
\usetikzlibrary{positioning, arrows.meta, shapes.geometric, trees}
\usepackage{pgfplots}
\pgfplotsset{compat=1.18}
\usepackage{svg}
\usepackage[most]{tcolorbox}
\usepackage{graphbox}
\usepackage{mathtools}
\usepackage{fancybox}
\usepackage{makecell}
\usepackage{multirow}
\usepackage{booktabs}
\usepackage{hyperref}
\hypersetup{hidelinks}
\usepackage{cleveref}
\usepackage[normalem]{ulem} \usepackage{listings}
\usepackage{caption}
\usepackage{subcaption}
\usepackage{longtable}
\usepackage{pifont}\usepackage{enumitem}
\usepackage{subfloat}
\usepackage[latin1]{inputenx}
\usepackage[T1]{fontenc}
\usepackage{soul}
\usepackage{courier}
\usepackage[english]{babel}

\usepackage[framemethod=TikZ]{mdframed}
\newenvironment{Answer}[1][]{
    \ifstrempty{#1}{
        \mdfsetup{
            frametitle={
                \tikz[baseline=(current bounding box.east), outer sep=0pt]
                \node[line width=0pt, anchor=east, rectangle, draw=white, fill=white]
            ;}
        }
    }{
        \mdfsetup{
            frametitle={
                \tikz[baseline=(current bounding box.east), outer sep=0pt]
                \node[anchor=east, rectangle, draw=white, fill=white]
                {\strut #1};
            }
        }
    }
    \mdfsetup{innertopmargin=-5pt, linecolor=black,
            linewidth=0.5pt, topline=true,
            frametitleaboveskip=\dimexpr-\ht\strutbox\relax,}
    \begin{mdframed}[]\relax
}{\end{mdframed}}

\newcommand\phase[1]{\tikz[baseline=(X.base)]\node [draw=black, fill=white, thick, rectangle, inner sep=2pt, rounded corners=2pt](X){\color{black}\textbf{#1}};}

\newboolean{showcomments}
\setboolean{showcomments}{true}
\ifthenelse{\boolean{showcomments}}{
    \newcommand{\nb}[2]{
        \fcolorbox{black}{yellow}{\bfseries \sffamily \scriptsize#1}
        {\sf \small $\blacktriangleright$ \textit{#2} $\blacktriangleleft$}
    }
    
}{
    \newcommand{\nb}[2]{}
    
}

\newcommand{\fga}{FGA\xspace}
\newcommand{\efga}{EFGA\xspace}
\newcommand{\fgp}{CBP\xspace}

\newcommand{\rival}{RIVAL10\xspace}
\newcommand{\vgg}{VGG-19\xspace}

\begin{document}

\title{Engineering Resource-constrained Software Systems with DNN Components: a Concept-based Pruning Approach}

\author{Federico Formica}
\authornote{Both authors contributed equally to this research.}
\email{formicaf@mcmaster.ca}
\orcid{0000-0002-3033-7371}
\affiliation{\institution{McMaster University}
    \city{Hamilton}
    \state{ON}
    \country{Canada}
}

\author{Andrea Rota}
\authornotemark[1]
\email{a.rota51@studenti.unibg.it}
\orcid{0009-0008-1648-4130}
\affiliation{\institution{University of Bergamo}
    \city{Bergamo}
    \country{Italy}
}

\author{Aurora Francesca Zanenga}
\email{aurora.zanenga@unibg.it}
\orcid{0009-0008-0655-8335}
\affiliation{\institution{University of Bergamo}
    \city{Bergamo}
    \country{Italy}
}

\author{Andrea Bombarda}
\email{andrea.bombarda@unibg.it}
\orcid{0000-0003-4244-9319}
\affiliation{\institution{University of Bergamo}
    \city{Bergamo}
    \country{Italy}
}

\author{Mark Lawford}
\email{lawford@mcmaster.ca}
\orcid{0000-0003-3161-2176}
\affiliation{\institution{McMaster University}
    \city{Hamilton}
    \state{ON}
    \country{Canada}
}

\author{Lionel C. Briand}
\email{lbriand@uottawa.ca}
\orcid{0000-0002-1393-1010}
\affiliation{\institution{University of Ottawa}
    \city{Ottawa}
    \state{ON}
    \country{Canada}
}
\affiliation{\institution{Lero Centre, University of Limerick}
    \city{Limerick}
    \country{Ireland}
}

\author{Claudio Menghi}
\email{claudio.menghi@unibg.it}
\orcid{0000-0001-5303-8481}
\affiliation{\institution{University of Bergamo}
    \city{Bergamo}
    \country{Italy}
}
\affiliation{\institution{McMaster University}
    \city{Hamilton}
    \state{ON}
    \country{Canada}
}

\renewcommand{\shortauthors}{Formica et al.}

\begin{abstract}
Deep Neural Networks (DNNs) are widely used by engineers to solve difficult problems that require predictive modeling from data.
However, these models are often massive, with millions or billions of parameters, and require substantial computational power, RAM, and storage.
This becomes a limitation in practical scenarios where strict size and resource constraints must be respected.
In this paper, we present a novel concept-based pruning technique for DNNs that guides pruning decisions using human-interpretable concepts, such as features, colors, and classes. 
This is particularly important in a software engineering context, as DNNs are integrated into systems and must be pruned according to specific system requirements.
Our concept-based pruning solution analyzes neuron activations to identify important neurons from a system requirements viewpoint and uses this information to guide the DNN pruning.
We assess our solution using the VGG-19 network and a dataset of 26'384 RGB images, focusing on its ability to produce small, effective pruned DNNs and on the computational complexity and performance of these pruned DNNs.
We also analyzed the pruning efficiency of our solution and compared alternative configurations.
Our results show that concept-based pruning efficiently generates much smaller, effective pruned DNNs.
Pruning greatly improves the computational efficiency and performance of DNNs, properties that are particularly useful for practical applications with stringent memory and computational time constraints.
Finally, alternative configuration options enable engineers to identify trade-offs adapted to different practical situations.
 \end{abstract}


\maketitle

\section{Introduction}
\label{sec:intro}
Software engineering has been deeply disrupted by AI~\cite{Terragni2025,uchitel2024scoping,MartinezFernandez2022,Fan2023,Liang2024}. 
While software developers were traditionally focused on writing software code, they now rely on AI solutions for many functionalities, leading to many AI-based software components being integrated into complex software systems~\cite{uchitel2024scoping}.
For example, Deep Neural Networks (DNNs) have achieved strong performance across domains such as computer vision~\cite{yolov8,DosovitskiyB0WZ21}, medical imaging~\cite{janowczyk2016}, and natural language processing~\cite{DevlinCLT19,BahdanauCB14} and are used in large software systems, such as Google Search~\cite{GoogleSearch}, Tesla~\cite{TeslaAI2023} autonomous driving system, virtual assistants (e.g., Apple Siri~\cite{capes17_interspeech}), and many others.

Unlike traditional development activities, in which software engineers had to write code---and carefully analyze its performance---to solve practical problems, engineers now must manage the increasing size of AI models~\cite{Amershi2019,Kriens2022} and their impact on system performance and resource usage. 
Indeed, DNN architectures have grown substantially in size and computational demand, often requiring powerful hardware, considerable memory, and bandwidth to operate effectively.
For example, \vgg~\cite{VGG_19} and ModernBERT$_{base}$~\cite{warner2025smarter} have 144M and 149M parameters, respectively, with sizes of 575MB and 599MB.
Therefore, in many practical situations, the size of DNNs does not align with the resource-constrained environments in which intelligent systems are increasingly expected to operate, thus hampering system design.

This is, for example, the case of edge devices, which are hardware components (e.g., sensors) that operate at the boundary of a network~\cite{Bombarda2025}.
These devices can process data locally (near the source) rather than relying on a centralized infrastructure.
This computing paradigm is particularly beneficial, as cloud-based inference imposes high latency, bandwidth constraints, and security and privacy issues, limiting its usability in time-sensitive domains such as autonomous driving, industrial monitoring, and medical wearables.
With the rise of Edge AI, where inference is executed directly on end-user devices, such as smartphones, wearables, Internet-of-Things (IoT) sensors, and embedded systems, deploying these large models becomes challenging~\cite{Meuser2024}.
By keeping data on-device, edge processing minimizes exposure to data breaches and allows AI systems to operate even when connectivity is intermittent. 
However, despite its advantages, it also introduces new constraints on power consumption, memory, and computational load~\cite{Ngo2025}.

A concrete example of such a resource-constrained environment, where privacy and real-time computing are mandatory requirements, is ECG Abnormality Detection. Models such as ConvLSTM2D-liquid time-constant and ConvLSTM2D-closed-form continuous-time~\cite{Huang2024} have been developed to run on the STM32F746G microcontroller (216 MHz CPU, 340 KB of RAM)~\cite{STMicroelectronics}, illustrating the hardware, privacy, and time-sensitive constraints under which modern intelligent systems must operate.
Similarly, small Recurrent NNs (RNNs) must be deployed on hearing-aid hardware, which is battery-powered and runs on resource-constrained microcontrollers with limited memory capacity~\cite{Fedorov2020}.
Another example comes from the aerospace domain, where embedded computing platforms must operate under strict energy, memory, and reliability constraints. For instance, the Raspberry Pi Zero W (512 MB RAM) has been employed as a flight computer and sensor control unit in CubeSat missions such as GASPACS~\cite{GASPACS,Raspberry}.

Model compression techniques have emerged as a practical solution to this problem~\cite{Li2023,Dantas2024,Cheng2024}.
For example, a recent study~\cite{li2021deep} evaluated seven combinations of model compression techniques for online fault detection in the Tennessee Eastman Chemical process. Apple~\cite{AppleVoiceTrigger,AppleFoundationModels} uses model compression techniques to enable DNNs to run on their devices.
Amazon Alexa employs model compression techniques to reduce the size and computational cost of speech and language models~\cite{AmazonAlexaOnDevice}.
In existing techniques, model pruning typically removes components of a neural network while limiting the accuracy loss of the DNN.
Prior work shows that pruning larger models can outperform training smaller dense models directly~\cite{zhu2017prune,li2020train}, further motivating pruning-based workflows.

Although model pruning has received considerable attention~\cite{Cheng2024,He2024,blalock2020}, 
existing pruning criteria are predominantly numerical. Magnitude-based approaches, for instance, assume that weights with small absolute values are expendable. 
But such criteria operate without any understanding of what a component actually does and the system context in which the DNN is integrated, and may therefore be suboptimal from a system viewpoint.
In contrast, we advocate using concepts relevant to the target system requirements to guide DNN pruning, aiming to tailor the DNN to the system's needs and thereby reduce its size without significantly affecting its accuracy in the system context.

In this paper, we propose \emph{concept-based pruning} (\fgp), a process guided by the selection of relevant concepts in a system context. 
Concepts are human-interpretable entities that can be extracted from the system requirements.
They can represent classes or their attributes (a.k.a., feature labels~\cite{Gopinath_2023}).
For example, a DNN integrated into a pedestrian avoidance system should be pruned to focus on pedestrian characteristics (e.g., direction and speed) and on the class indicating their presence in the vehicle's field of view.
Another example from a recent work~\cite{formica2026ensemblesbasedfeatureguidedanalysis} considered the values of the digits from the MNIST dataset~\cite{Lecun1998} as their classes, and the presence of circles and lines within those digits as feature labels.

Concept-based pruning differs from standard magnitude-based pruning because it enables software engineers to guide pruning more effectively based on concepts that can be derived from the requirements of the system in which the DNN is to be integrated. 
Indeed, using concepts to drive the pruning enables the removal of high-value weights that do not contribute to the model's decision-making relevant to the system.

We implemented an instance of our general concept-based approach that uses Feature-Guided Analysis (\fga)~\cite{Gopinath_2023} and its ensemble extension (\efga)~\cite{formica2026ensemblesbasedfeatureguidedanalysis} to identify relevant neurons of a DNN for a set of concepts. 
We then use the Torch-Pruning tool~\cite{fang2023depgraph} to remove neurons that are not useful for detecting the presence of high-level concepts selected for their relevance in a system context.

We evaluated our solution using the \vgg NN architecture and the \rival dataset. We considered publicly available weights for \vgg, pretrained on ImageNet dataset. We selected the \rival dataset, a subset of ImageNet, and identified 10 relevant concepts corresponding to the classes present in \rival.
We assessed \fgp in terms of its ability to generate small and effective pruned networks (RQ1), its capability to improve the computational complexity and performance of a DNN (RQ2), and its efficiency in producing pruned DNNs (RQ3).
We also compared the rules generated by different \fga configuration options (RQ4).

Our results show that \fgp is effective in generating compact pruned DNNs while maintaining acceptable predictive performance.
\fgp can significantly reduce the size of the network layers under analysis and improve network performance.
Furthermore, it can generate the pruned DNN in a practical time. 
Finally, different configuration options offer alternative trade-offs between network size and accuracy that can be beneficial depending on the application domain.

To summarize, the contributions of this paper are:
\begin{itemize}
    \item A novel concept-based pruning framework (\fgp) and its implementation (\Cref{sec:pruning}), targeting the integration of DNNs into a specific system, where the pruned DNNs must satisfy its requirements (e.g., a subset of relevant classes).
    \item An extensive empirical evaluation addressing \fgp's effectiveness, computational impact, efficiency, and sensitivity to misclassified samples (\Cref{sec:evaluation}).
\end{itemize}

Our paper is organized as follows.
\Cref{sec:pruning} presents our concept-based pruning framework and its implementation.
\Cref{sec:evaluation} evaluates our contribution.
\Cref{sec:discussion} discusses our results and threats to validity.
\Cref{sec:related} summarizes related work.
\Cref{sec:conclusion} presents our conclusions.

\section{Concept-Based Pruning}
\label{sec:pruning}
We first present our concept-based pruning framework (\Cref{sec:framework}) and a proposed implementation (\Cref{sec:implementation}).

\subsection{Overview}
\label{sec:framework}
\begin{figure}[t]
    \centering
    \begin{tikzpicture}[
    >=Latex,
    font=\small,
    block/.style={
        draw,
        rounded corners,
        align=center,
        thick,
    },
    terminator/.style={
        rounded corners,
        align=center,
    }
]

\node[terminator] (start) {Dataset};
\node[terminator, above=0.44cm of start] (start_2) {Concepts};

\node[block, right=0.5cm of start] (methods)
{\phase{1} Neurons\\Identifier};

\node[block, right=1.5cm of methods] (prune)
{\phase{2} Pruner};

\node[terminator, right=0.5cm of prune] (pruned) {Pruned DNN};

\node[above right=0.25cm and 0.25cm of methods] (dnn) {DNN};

\draw[->] (start) -- (methods);
\draw[->] (start_2) -| ([xshift=-0.25cm]methods);

\draw[->] (methods) -- node[below, align=center] 
{Neurons} (prune);

\draw[->] (prune) -- (pruned);

\draw[->] (dnn) -| ([xshift=0.25cm]methods);
\draw[->] (dnn) -| (prune);
\draw[->] (pruned.north) -- ++(0,1.25) coordinate (top)
          -- node[above] {\textit{substitute}} (dnn.north |- top)
          -- (dnn.north);

\end{tikzpicture}     \caption{Concept-Based Pruning.}
    \label{fig:pruning-algorithm}
    \Description{Flowchart representing the proposed approach: Concept-Based Pruning.
    The flowchart has two main phases: (i) the Neuron Identifier and (ii) the Pruner.
    The Neuron Identifier takes as input a Dataset, a set of Concepts, and a Deep Neural Network, and outputs a list of Neurons.
    The Pruner takes as input the list of Neurons and the DNN and returns the pruned DNN.
    Finally, there is a feedback loop in which the pruned DNN replaces the original DNN, allowing the pruning process to be repeated iteratively.}
\end{figure}
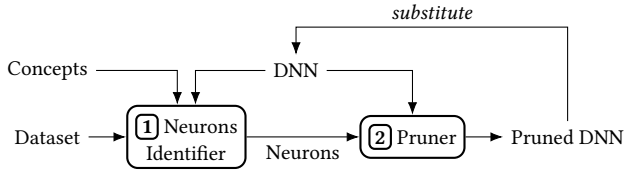

\begin{figure}[t]
   \centering
   \subfloat[][A horse image labeled Equine (ImageNet: Sorrel).]{
       \includegraphics[width = 0.45\linewidth]{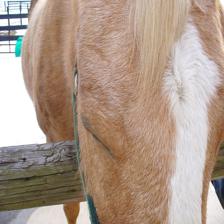}
       \label{fig:rival-example-1}
   }
   \hfill
   \subfloat[][An airplane image labeled Plane (ImageNet: Airliner).]{
       \includegraphics[width = 0.45\linewidth]{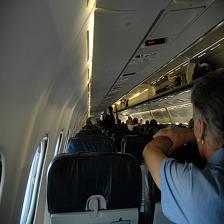}
       \label{fig:rival-example-2}
   }\\
   \caption{Examples images from the \rival benchmark.}
   \label{fig:rival-example}
   \Description{The figure contains two examples of images from the \rival benchmark.
   The first image shows a horse's muzzle, taken from a short distance. The horse's head covers the rest of the body. The image is labeled with the Equine class.
   The second image shows the interior of an airplane taken from a window seat. The image is labeled with the Plane class.}
\end{figure}

\Cref{fig:pruning-algorithm} introduces our \emph{concept-based pruning} (\fgp) framework.
Concept-based pruning identifies the \emph{neurons} used by the network to produce its outputs.
Then, it uses this information to guide the pruning task.
Our framework takes as input a trained DNN and a dataset of images relevant to the system (selected from the training dataset or a different one).
Each image is associated with concepts: class and feature labels.
For instance, \Cref{fig:rival-example-1} and \Cref{fig:rival-example-2} present two images from the \rival benchmark: the former belongs to the class equine and is characterized by the features mane, hairy, and patterned, whereas the latter belongs to the class plane and is characterized by the features metallic, long, and tall.

\fgp consists of two components: \emph{Neurons Identifier} (\phase{1}) and \emph{Pruner} (\phase{2}).

The \emph{Neurons Identifier} (\phase{1}) extracts the neurons responsible for recognizing specific concepts for a given DNN and dataset.
The framework from \Cref{fig:pruning-algorithm} is generic.
Although our implementation is generic, the \emph{Neurons Identifier} component can be implemented differently depending on the application domain, the pruning goal, and the dataset.
\Cref{sec:implementation} presents a possible implementation of the \emph{Neurons Identifier} component.

The \emph{Pruner} (\phase{2}) receives the set of neurons identified by the \emph{Neurons Identifier} and the DNN, and prunes the network by removing all neurons that are not in this set.
As the \emph{Neurons Identifier}, the \emph{Pruner} component can also be implemented differently.
Our framework enables engineers to implement various pruning strategies according to their objectives, such as removing or zeroing weights, or deleting channels and filters from the model.

Given the identification and pruning strategies, our approach can preserve the model's original functionality (Original Task) or specialize the network on a subset of concepts (Transfer Pruning).
The former objective aims to reduce the size of a DNN, whereas the latter aims to produce highly specialized models for specific tasks. For example, a security camera model may be pruned to recognize only people while discarding concepts related to animals or vehicles.

To increase the pruning level, the pruned DNN can be fed back into the \emph{Neurons Identifier} (\phase{1}). The process is then repeated until a user-defined stopping condition is met. For example, we propose the following stopping criteria: (i)~no progress compared to the previous iteration (i.e., no neurons were removed), (ii)~the maximum number of iterations is reached, (iii)~a target model size is achieved, or (iv)~a minimum acceptable accuracy (or precision/recall) is reached.

\begin{figure}[t]
    \centering
    \begin{tikzpicture}[
    >=Latex,
    font=\small,
    level 1/.style={sibling distance=4.5cm},
    level 2/.style={sibling distance=2cm},
    level distance=1.2cm,
    edge from parent/.style={draw},
    every node/.style={align=center}
]

\node {$N_{1,12}$}
child {
    node {$N_{1,543}$}
        child {
            node {\emph{equine-present}\\(984,0)}
            edge from parent node[left] {$\le 3.21$}
        }
        child {
            node {\emph{equine-absent}\\(0,1238)}
            edge from parent node[right] {$> 3.21$}
        }
    edge from parent node[left] {$\le 1.65$}
}
child {
    node {$N_{1,1843}$}
        child {
            node {\emph{equine-present}\\(4290,0)}
            edge from parent node[left] {$\le 2.93$}
        }
        child {
            node {\emph{equine-absent}\\(0,244)}
            edge from parent node[right] {$> 2.93$}
        }
    edge from parent node[right] {$> 1.65$}
};

\end{tikzpicture}     \caption{Example of decision tree extracted by \fga.}
    \label{fig:implementation-decision-tree}
    \Description{Example of a decision tree extracted by FGA.
    The decision tree in question is complete, with three layers and four leaf nodes.
    Each non-leaf node contains the identifier of a neuron, and each branch contains a condition of the activation layer of that neuron.
    Each leaf node contains a clause specifying whether the concept is present or not, and the number of elements in the dataset that are represented by that leaf node.}
\end{figure}
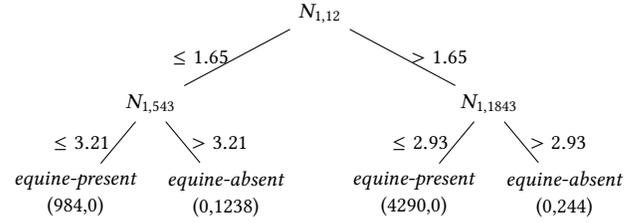

\subsection{Implementation}
\label{sec:implementation}

We implemented the components of our solution (\emph{Neurons Identifier} --- \phase{1} and the \emph{Pruner} --- \phase{2})  as follows. 
We remark that this is one possible implementation, and alternative components realizing the functionalities outlined in \Cref{sec:framework} can be used.

\emph{Neurons Identifier} (\phase{1}).  We propose two alternative components: one based on Feature-Guided Analysis (\fga)~\cite{Gopinath_2023} and one based on its extension, Ensemble-based Feature-Guided Analysis (\efga)~\cite{formica2026ensemblesbasedfeatureguidedanalysis}.
We considered two alternative components because, in our evaluation, we will assess how sensitive \fgp is to different configuration options, such as different implementations of the \emph{Neurons Identifier} (\phase{1}) component (see \Cref{sec:evaluation}).
We selected \fga because (a)~it can extract neurons related to specific concepts\footnote{An \fga concept can represent a class or an input feature.}, (b)~the results from two case studies from the aerospace (TaxiNet~\cite{Frew2004,Beland2020}) and the automotive domain (YOLOv4-Tiny~\cite{Caesar2019nuScenesAM}) confirm its effectiveness, and (c)~the results have been confirmed by a recent replication study~\cite{FGA_Replication}.
We provide a brief introduction to \fga, although it is not necessary to understand in detail \fga to understand our contribution.
A precise description is out of scope, and the interested reader can refer to the corresponding publication~\cite{Gopinath_2023}.
\fga considers a DNN, a dataset, and a set of concepts of interest.
For each image from the dataset, \fga extracts the activation values of all neurons and a set of labels indicating the presence or absence of the concept of interest.
Then, for every concept, \fga extracts a decision tree that defines conditions on neuron activation values entailing the presence or absence of that concept.
Notice that \fga first extracts a decision tree, and then converts it into decision rules. 
Alternative implementations can directly extract decision rules and consider more sophisticated rule-based algorithms (e.g., RuleFit~\cite{molnar2025}).
\Cref{fig:implementation-decision-tree} shows an illustrative example of a decision tree computed by \fga related to the concept ``equine''. Each node refers to a neuron $\text{N}_{x,y}$ (layer $x$, neuron $y$), and edges are labeled with conditions on activation values. Leaf nodes are associated with a tuple $(a,b)$, where $a$ and $b$ are the number of inputs labeled as concept-present and concept-absent, respectively.
A leaf node is considered pure when it is labeled as concept-present and $b = 0$, or when it is labeled as concept-absent and $a = 0$.
For example, the leftmost leaf node from \Cref{fig:implementation-decision-tree} is pure since the \emph{equine} concept is present and $b = 0$.
A path from the root to a pure leaf defines a decision rule in the form \textbf{pre}~$\rightarrow$~\textbf{post}, where \textbf{pre} is a conjunction of neuron activation conditions and \textbf{post} indicates the presence or absence of the concept. For example, from \Cref{fig:implementation-decision-tree}, \fga extracts:
\begin{itemize}
    \item[] $(\text{N}_{1,12} \leq 1.65 \land \text{N}_{1,543} \leq 3.21) \rightarrow \emph{equine-present}$.
    \item[] $(\text{N}_{1,12} > 1.65 \land \text{N}_{1,1843} \leq 2.93) \rightarrow \emph{equine-present}$.
\end{itemize}
\fga considers only pure leaves; all other nodes are not considered for the creation of a rule. 
Unlike the original implementation of \fga~\cite{Gopinath_2023}, our version of \fga for \fgp returns the complete set of identified rules, preventing \fgp from being overly aggressive in pruning.
Indeed, by considering the original \fga implementation, which selects only one rule per concept, we would prune large parts of the DNN, significantly affecting its performance.
Our solution extracts neurons from the preconditions of these rules to identify which neurons contribute to the correct model output when active and which do not, and therefore could be removed.

Our second implementation for the component relies on Ensemble Feature-Guided Analysis (\efga)~\cite{formica2026ensemblesbasedfeatureguidedanalysis}.
We selected \efga because it improves the results from \fga by increasing the recall of the rules returned by \fga.
Specifically, \efga returns a rule that combines more rules from \fga into a single rule according to a performance metric.
\efga offers alternative options to aggregate rules.
For each concept and layer, the \textbf{TOP(N)} option aggregates the $N$ rules with the highest training recall into a single rule, 
the \textbf{REC(X)} option aggregates rules into a single one until a cumulative training recall above $X\%$ is reached, the \textbf{AVG} option aggregates rules with a training recall above the average recall of extracted rules into a single rule. 
Intuitively, \textbf{TOP(N)} requires engineers to decide a priori (N) the number of rules to be aggregated, \textbf{REC(X)} requires engineers to set a desired threshold on the recall (X), while \textbf{AVG} aggregates rules that have a training recall value above average. 
These diverse options may be valuable in different contexts. 
For example, by setting a recall threshold, engineers can control the number of relevant concepts the rule successfully detects. 
However, unlike \textbf{TOP(N)}, \textbf{REC(X)} does not impose a bound on the number of rules that can be aggregated, which is relevant when having succinct rules is of interest.

To implement \fgp, we adapted the publicly available code of \fga and \efga.
We migrated from Keras 2 to PyTorch to enable the use of our pruning implementation.
Note that the current implementation of \fga and \efga employs decision trees to represent rules. However, in future work, we may explore implementing more advanced approaches, such as RuleFit~\cite{Friedman2008}, which has been shown to outperform decision trees in many cases.

\emph{Pruner} (\phase{2}). We used the ``Torch-Pruning'' tool~\cite{fang2023depgraph}.
It is a Python library designed for pruning DNNs in PyTorch. 
Torch-Pruning models dependencies between layers explicitly, comprehensively groups coupled parameters, and performs the pruning accordingly.
This tool provides a set of low-level pruning utilities that remove specific structural components of a neural network while maintaining architectural consistency. 
It is widely used, with 510k downloads according to PyPI~\cite{torch_pruning_pypi}, and has been used in other works~\cite{Li_2023_CVPR,fang2024isomorphic}.
In this implementation, our pruning is structured: we remove neurons from fully connected layers (and their associated connections), rather than zeroing individual weights.

Our implementation offers different configuration options.
First, it enables engineers to configure how to treat samples misclassified by the DNN, i.e., input samples for which the network's predicted label differs from the correct one.
Indeed, our implementation enables engineers to decide whether \fgp treats misclassified samples as correct and includes them in the decision tree computation, or discards them. 
Although we generally believe these samples should be discarded, we support their use since, in the original implementation of \fga~\cite{Gopinath_2023}, misclassified samples are used to build the decision trees.\footnote{Note that \fga was initially used to explain the behavior of a DNN.}
Second, it enables engineers to select \fga or \efga for neuron extraction.
\efga enables considering ensembles to increase the recall of the rule returned by \fga.
Therefore, they enable the extraction of rules involving a larger number of neurons, making the pruning less aggressive.
Finally, if \efga is selected, engineers can consider alternative aggregation strategies (TOP(N), REC(X), and AVG).

Our default configuration does not account for samples misclassified by the DNN and uses \fga to compute the pruned DNN.
We chose this configuration to preserve neurons that contribute to correct predictions, rather than retaining those associated with incorrect predictions. Furthermore, \fga is more conservative because it uses all extracted rules, whereas \efga selects only a subset based on the chosen aggregation strategy (i.e., TOP(N), REC(X), and AVG). Different configurations are discussed in \Cref{sec:rq4}.

 \section{Evaluation}
\label{sec:evaluation}

We evaluated \fgp by considering the following research questions (RQs):
\begin{enumerate}[start=1,label={\bfseries RQ\arabic*:}]
    \item What is the \emph{trade-off} in terms of size reduction and accuracy obtained with \fgp? (\Cref{sec:rq1})
    \item How effective is \fgp in \emph{improving computational complexity and performance}? (\Cref{sec:rq2}) 
\end{enumerate}
These two research questions assess how our concept-based pruning solution reduces the size of the DNN and improves its performance compared to the original DNN.
The goal is to assess whether considering human-interpretable concepts (derived from requirements) expressed by class and feature labels enables the creation of an effective pruned DNN.   
\begin{enumerate}[start=3,label={\bfseries RQ\arabic*:}]
    \item How efficient is \fgp in \emph{reducing the size} of DNNs? (\Cref{sec:rq3})
\end{enumerate}
This research question evaluates the efficiency of \fgp.
The goal is to assess if \fgp can prune large DNNs in practical time.
\begin{enumerate}[start=4,label={\bfseries RQ\arabic*:}]
\item How do the rules generated by different \fga \emph{configuration options} compare? (\Cref{sec:rq4})
\end{enumerate}
This research question evaluates how different configuration options affect the length and completeness of the rules extracted by \fgp, and how this impacts the effectiveness of the \fgp approach.

\emph{Benchmark}. To answer our research questions, we considered the Rich Visual Attributes with Localization (\rival)~\cite{Rival_10} dataset as our benchmark. 
The \rival dataset includes the CIFAR-10 classes~\cite{cifar10} (i.e., ``bird'', ``car'', ``cat'', ``deer'', ``dog'', ``equine'', ``frog'', ``plane'', ``ship'', and ``truck'') representing high-level concepts that can be extracted from requirements.
It does so by combining two ImageNet-1k~\cite{ImageNet} labels (e.g., combining different types of dogs into class dog) for each \rival class, resulting in a subset of ImageNet samples. This emulates cases in which a system requires a DNN for a subset of classes and coarser-grained classification. 
\Cref{fig:rival-example} shows two example images from this dataset. The dataset consists of 26'384 RGB images (21'098 for training and 5'286 for testing), each of size 224 $\times$ 224 pixels, and it is balanced: each class has between 2523 and 2667 samples.
\rival is widely used as an ImageNet-derived benchmark for robustness, explainability, and concept-based analysis~\cite{Moayeri2022}, and adopted in several recent works~\cite{AhmadiRKS23,selvaraj2024improving,mangal2024concept,Prophecy_2025,Santos2024}.
This benchmark is particularly suitable for assessing our concept-based pruning, as it was used to evaluate a logical specification language for concept-based requirements~\cite{mangal2024concept}, thereby demonstrating the central role of concepts in this benchmark.

\emph{Study Subject}. Our study subject is the  \vgg~\cite{VGG_19} DNN architecture.
We used publicly available pretrained ImageNet weights from the PyTorch framework~\cite{Pytorch_vgg_19} for this DNN. 
The choice of our study subject is motivated by two factors. 
First, the pretrained weights are well-suited for \rival, since the latter is a subset of ImageNet. 
Second, \vgg has demonstrated a strong generalization across a wide range of classification tasks, including plant disease detection and fruit detection for smart agriculture~\cite{Nguyen2022,Sajid2025}, medical image diagnosis~\cite{ALSHMRANI2023,Dey2021}, violence detection in video~\cite{Negre2026}, and industrial fault diagnosis~\cite{Barrera2023}, confirming its robustness as a feature extractor. 
Notably, \vgg is frequently employed not only as a standalone classifier but also as a backbone within larger pipelines and with task-specific architectural modifications, further attesting to its flexibility~\cite{Negre2026,Dey2021}.

The architecture of \vgg consists of 16 convolutional layers and 3 fully connected layers, arranged sequentially. 
\Cref{tab:benchmark-vgg19} details the number of outputs and parameters for each layer.
In this work, we analyzed the first and second fully connected layers (FC1 and FC2) because each contains 4096 neurons, and together they account for 86.06\% of the network's trainable parameters (including the weights of FC1 and FC2 and the input connections of FC3). FC3 was excluded from the analysis, as it is directly tied to the classification output. 
On the \rival dataset, this model achieves an accuracy of 84.79\%, with a precision of 90.74\% and a recall of 77.10\%.

\emph{Experimental Methodology}. To answer RQ1, RQ2, RQ3, and RQ4, we iteratively ran our concept-based pruning solution using the  \rival dataset and the entire set of generated rules. 
All experiments were executed on an Apple MacBook Pro equipped with an Apple M1 Pro processor and 16 GB of RAM.
We did not repeat the experiments because there are no stochastic elements.
We set the maximum number of pruning iterations to 100 (see \Cref{sec:pruning}) and save the pruned network after each iteration.
Depending on the specific research question, we considered different metrics.

\begin{table}[t]
\footnotesize
\centering
\caption{Architecture and number of parameters of \vgg.}
\label{tab:benchmark-vgg19}
\begin{tabular}{llr}
\toprule
\textbf{Layer} & \textbf{Output Size} & \textbf{\# Params} \\
\midrule

Input & 224 $\times$ 224 $\times$ 3 & 0 \\

\midrule
Conv1\_1 & 224 $\times$ 224 $\times$ 64 & 1'792 \\
Conv1\_2 & 224 $\times$ 224 $\times$ 64 & 36'928 \\
MaxPool  & 112 $\times$ 112 $\times$ 64 & 0 \\

\midrule
Conv2\_1 & 112 $\times$ 112 $\times$ 128 & 73'856 \\
Conv2\_2 & 112 $\times$ 112 $\times$ 128 & 147'584 \\
MaxPool & 56 $\times$ 56 $\times$ 128 & 0 \\

\midrule
Conv3\_1 & 56 $\times$ 56 $\times$ 256 & 295'168 \\
Conv3\_2 & 56 $\times$ 56 $\times$ 256 & 590'080 \\
Conv3\_3 & 56 $\times$ 56 $\times$ 256 & 590'080 \\
Conv3\_4 & 56 $\times$ 56 $\times$ 256 & 590'080 \\
MaxPool  & 28 $\times$ 28 $\times$ 256 & 0 \\

\midrule
Conv4\_1 & 28 $\times$ 28 $\times$ 512 & 1'180'160 \\
Conv4\_2 & 28 $\times$ 28 $\times$ 512 & 2'359'808 \\
Conv4\_3 & 28 $\times$ 28 $\times$ 512 & 2'359'808 \\
Conv4\_4 & 28 $\times$ 28 $\times$ 512 & 2'359'808 \\
MaxPool  & 14 $\times$ 14 $\times$ 512 & 0 \\

\midrule
Conv5\_1 & 14 $\times$ 14 $\times$ 512 & 2'359'808 \\
Conv5\_2 & 14 $\times$ 14 $\times$ 512 & 2'359'808 \\
Conv5\_3 & 14 $\times$ 14 $\times$ 512 & 2'359'808 \\
Conv5\_4 & 14 $\times$ 14 $\times$ 512 & 2'359'808 \\
MaxPool  & 7 $\times$ 7 $\times$ 512 & 0 \\

\midrule
FC1 & 4096 & 102'764'544 \\
FC2 & 4096 & 16'781'312 \\
FC3 & 1000 & 4'097'000 \\

\midrule
\textbf{Total} &  & \textbf{143'667'240} \\
\bottomrule
\end{tabular}
\end{table}

 \subsection{Effectiveness of the Pruned DNN (RQ1)}
\label{sec:rq1}
To assess the size reduction and the effectiveness of the pruned DNNs generated by \fgp, we considered the following metrics.

\emph{Metrics}.
We considered two categories of metrics: the first assesses the reduction in DNN size, and the second assesses its effectiveness.
For network reduction, we report the number of neurons of the pruned DNN in \textbf{FC1} and \textbf{FC2} (out of 4096 each), the total parameter count (\textbf{Params}), and the model size in megabytes (\textbf{Size}). \textbf{Params} denote the total number of trainable parameters of the model. 
\textbf{Size} corresponds to the file size in MB of the model saved via PyTorch's (\texttt{torch.save()}). 
For effectiveness, we consider four metrics: \textbf{accuracy} ($\nicefrac{TP + TN}{TP + FP + TN + FN}$), \textbf{precision} ($\nicefrac{TP}{TP + FP}$), \textbf{recall} ($\nicefrac{TP}{TP + FN}$), and \textbf{F1-score} ($\nicefrac{2TP}{2TP + FP + FN}$), where TP, FP, TN, and FN are defined as follows.
A True Positive (TP) denotes a correct prediction of the presence of a concept, while a False Positive (FP) identifies the concept as present when it is not. 
Similarly, a True Negative (TN) denotes a correct prediction of the absence of a concept, while a False Negative (FN) identifies the concept as absent when it is in fact present.

\begin{table*}[t]
\centering
\footnotesize
\caption{Size and effectiveness of the pruned DNN generated by \fgp.}
\label{tab:rq1-results}
\begin{tabular}{l | c c c c | c c c c}
\toprule
& \multicolumn{4}{c|}{\textbf{Size}} & \multicolumn{4}{c}{\textbf{Effectiveness}} \\
\textbf{Iteration} & \makecell{\textbf{FC1}} & \makecell{\textbf{FC2}} & \makecell{\textbf{Params}} & \makecell{\textbf{Size}} & \makecell{\textbf{Accuracy}} & \makecell{\textbf{Precision}} & \makecell{\textbf{Recall}} & \makecell{\textbf{F1-score}} \\
& \scriptsize{(neurons)} & \scriptsize{(neurons)} & \scriptsize{(M)} & \scriptsize{(MB)} & \scriptsize{(\%)} & \scriptsize{(\%)} & \scriptsize{(\%)} & \scriptsize{(\%)} \\
\midrule
\vgg        & 4096 & 4096 & 143.67 & 574.70 & 84.79 & 90.74 & 77.10 & 83.22 \\
1           & 2622 & 2357 & 94.35  & 377.42 & 81.71 & 90.73 & 74.31 & 81.48 \\
10          & 1241 & 1088 & 53.60  & 214.43 & 78.34 & 90.60 & 71.29 & 79.32 \\
20          & 1007 & 903  & 47.10  & 188.44 & 76.49 & 90.53 & 69.62 & 78.03 \\
30          & 903  & 823  & 44.25  & 177.02 & 76.83 & 90.38 & 69.93 & 78.13 \\
40          & 840  & 771  & 42.52  & 170.10 & 76.37 & 90.38 & 69.52 & 77.76 \\
50          & 804  & 747  & 41.55  & 166.21 & 75.48 & 90.42 & 68.71 & 77.15 \\
60          & 773  & 712  & 40.68  & 162.76 & 74.23 & 90.23 & 67.59 & 76.13 \\
70          & 744  & 676  & 39.87  & 159.51 & 75.24 & 90.12 & 68.50 & 76.75 \\
\bottomrule
\end{tabular}
\end{table*}

\begin{table*}[t]
\centering
\footnotesize
\caption{Computational cost and runtime analysis on \vgg's first two FC layers.}
\label{tab:rq2-results}
\begin{tabular}{l | c c | c c c | c c c }
\toprule
& \multicolumn{2}{c|}{\textbf{Active Neurons}} & \multicolumn{3}{c|}{\textbf{Computational Complexity}} & \multicolumn{3}{c}{\textbf{Inference Efficiency}} \\
\textbf{Iteration} & \makecell{\textbf{FC1}} & \makecell{\textbf{FC2}} & \makecell{\textbf{MACs FC1}} & \makecell{\textbf{MACs FC2}} & \makecell{\textbf{Total MACs}} & \makecell{\textbf{Latency}} & \makecell{\textbf{Std.}} & \makecell{\textbf{FPS}} \\
& \scriptsize{(neurons)} & \scriptsize{(neurons)} & \scriptsize{(M)} & \scriptsize{(M)} & \scriptsize{(G)} & \scriptsize{(ms)} & \scriptsize{($\pm$ms)} & \\
\midrule
\vgg        & 4096 & 4096 & 102.76 & 16.78 & 19.668 & 13.35 & 0.25 & 74.91 \\
1           & 2622 & 2357 & 65.78  & 6.18  & 19.619 & 12.27 & 0.19 & 81.50 \\
10          & 1241 & 1088 & 31.13  & 1.35  & 19.578 & 11.14 & 0.17 & 89.77 \\
20          & 1007 & 903  & 25.26  & 0.91  & 19.571 & 10.96 & 0.05 & 91.24 \\
30          & 903  & 823  & 22.65  & 0.74  & 19.569 & 10.88 & 0.06 & 91.91 \\
40          & 840  & 771  & 21.07  & 0.65  & 19.567 & 10.87 & 0.05 & 92.00 \\
50          & 804  & 747  & 20.17  & 0.60  & 19.566 & 10.86 & 0.17 & 92.08 \\
60          & 773  & 712  & 19.39  & 0.55  & 19.565 & 10.80 & 0.06 & 92.59 \\
70          & 744  & 676  & 18.67  & 0.50  & 19.564 & 10.79 & 0.04 & 92.68 \\
\bottomrule
\end{tabular}
\end{table*}

\emph{Results}. \Cref{tab:rq1-results} reports our results.
For conciseness, the table reports data collected every 10 iterations.
Our results do not include all 100 iterations, since after 70 iterations \fgp reaches a plateau: it stops removing neurons because all of them are included at least once in the extracted rules.

Our results show that \fgp is effective in generating small pruned networks.
Overall, after 70 iterations, our implementation reduces the model size from 574.70 MB to 159.51 MB (-72.24\%).
The effectiveness of the pruning process decreases with the number of iterations: \fgp prunes a higher number of neurons in early iterations. 
The number of pruned neurons per iteration decreases as the number of iterations increases.
For example, in the first iteration \fgp removes 1474 neurons (-35.99\%) from FC1 and 1739 neurons (-42.46\%) from FC2, reducing the total number of model parameters by 49,319,087 (-34.33\%), corresponding to -197.28 MB in model size, while in the last iteration \fgp removes 3 neurons (-0.07\%) from FC1 and 1 neuron (-0.02\%) from FC2, reducing the total number of model parameters by 79,043 (-0.06\%), corresponding to -0.32 MB in model size.

Our results on the effectiveness of the pruned DNNs show that the performance reduction offers an interesting trade-off across iterations.
For example, for the first iteration, the accuracy drops by -3.08\%, recall decreases by -2.79\%, F-1 score decreases by -1.74\%,  while precision remains stable (-0.01\%). 
This result suggests that a significant portion of the fully connected layers does not contribute to the final prediction for the concept present in the \rival dataset. 
However, these neurons may still contribute to classes outside the target concepts.
Furthermore, the DNN's effectiveness decreases with the number of iterations, reaching a plateau.
For example, between iterations 50, 60, and 70 in \Cref{tab:rq1-results}, effectiveness first slightly decreases (accuracy drops by -1.25\%, recall decreases by -1.12\%, F-1 score decreases by -1.02\%, while precision remains stable -0.19\%) and then slightly increases (accuracy improves by 1.01\%, recall increases by 0.91\%, F-1 score increases by 0.62\%, while precision remains stable -0.11\%). 
Overall precision remains substantially unchanged (decreasing only from 90.74\% to 90.12\%) even after 70 iterations, accuracy drops by -9.55\%, recall decreases by -8.60\%, and F-1 score decreases by -6.47\%.
This indicates that \fgp makes the network more specialized: while the pruned model becomes much smaller, its classification quality and reliability remain high, thus offering interesting trade-offs across iterations.

These results highlight that the engineer should choose a practical trade-off (an early-exit strategy) based on the application's requirements. 
If memory is the main constraint, pruning can be pushed further; if a minimum predictive quality is required, pruning should stop earlier. 
For example, using the practical scenario of a Raspberry Pi Zero W (512 MB RAM) discussed in \Cref{sec:intro}, an engineer could stop at the second iteration, where the model size is 315.44 MB (down from 574.70 MB) while the accuracy remains 80.65\%.
In this way, \fgp can be configured as a requirement-driven process rather than a fixed pruning schedule.

\begin{Answer}[RQ1 --- Effectiveness of the Pruned DNN]
The default configuration of \fgp is effective at generating compact DNNs: after 70 iterations, it removed 3352 neurons (-81.84\%) from FC1 and 3420 neurons (-83.50\%) from FC2, reducing the total number of parameters by 103,796,020 (-72.25\%), corresponding to a -415.18 MB reduction in model size.
The resulting reduction of the effectiveness of the DNN offers an interesting trade-off: 
the accuracy drops by -9.55\%, recall decreases by -8.60\%, F-1 score decreases by -6.47\%,  precision remains stable (-0.62\%). 
Engineers can configure \fgp to obtain different trade-offs between pruning and accuracy. 
\end{Answer} \subsection{Complexity and Efficiency of the Pruned DNN (RQ2)}
\label{sec:rq2}
To assess the impact of \fgp on computational complexity and inference efficiency of a DNN, we considered the following metrics.

\emph{Metrics}.
To quantify computational complexity, we aim to determine the number of operations a DNN requires to process an input. 
For this reason, we consider the number of Multiply-Accumulate operations (MACs) of the DNN, which measures the number of multiply-accumulate operations required for one forward pass.
MACs are widely used as a proxy for DNN complexity because the majority of DNN computation consists of linear algebra operations such as matrix multiplications and convolutions, which decompose into MACs~\cite{Saraswathy2025}.
We focus on the MACs derived from FC1 and FC2 since we configure \fgp to prune these two layers.

Regarding efficiency, we considered the  \textbf{Latency} (the time the DNN takes to produce an output from a single input), reported in milliseconds (ms) as the mean and standard deviation (\textbf{Std.}) per image over 100 runs; and Frames Per Second (\textbf{FPS}), computed as $1000 / \text{latency}$ when latency is expressed in milliseconds

\emph{Results}.
\Cref{tab:rq2-results} shows the changes in computational complexity and inference efficiency over the pruning iterations.

The computational complexity decreases with the number of iterations. 
\fgp achieves a greater reduction in MACs during early iterations, with such reduction per iteration decreasing progressively thereafter.
For example, in the first iteration of \fgp, the MAC reduction is -35.99\% for FC1 and -63.17\% for FC2, whereas between iterations 69 and 70, the additional reduction is only -0.07\% for FC1 and -0.02\% for FC2.
Overall, for the FC1 and FC2 layers, the MAC saving is significant: It reaches -81.83\% for FC1 and -97.02\% for FC2 by the final iteration.

The results from \Cref{tab:rq2-results} show that reducing the number of neurons leads to much better performance concerning Latency and FPS.
The inference efficiency of the pruned DNN improves with the number of iterations. 
The greatest gains in Latency and FPS are achieved during early iterations, with improvements diminishing steadily as pruning progresses.
After the first iteration, the model reduces latency by 1.08 ms, corresponding to a gain of 6.59 FPS (+8.80\%), thanks to a reduction of 3213 neurons. In the last iteration, the model's performance increases to 92.68 FPS (+23.73\%), resulting in a latency decrease of 2.56 ms (-19.18\%). 
Note that the time savings are particularly significant given that most of the computational complexity (99.39\%) comes from the convolutional layers, while FC1 (0.52\%) and FC2 (0.09\%) account for only a small part.

\begin{Answer}[RQ2 --- Computational Complexity and Performance]
\fgp significantly reduces the MACs from the FC1 (up to -81.83\%) and FC2 (up to -97.02\% ) layers and  latency (up to -19.18\%). It also increases FPS (up to +23.73\%).
\end{Answer}

\subsection{Efficiency of \fgp (RQ3)}
\label{sec:rq3}
To assess the efficiency of \fgp, we considered the following metrics.

\emph{Metrics}. 
We recorded the total time required by our algorithm (\textbf{Total}) and the time required by each of its two phases: The \emph{Neurons Identifier} (\phase{1}) and the \emph{Pruner} (\phase{2}).
We compute the sum of the times required by each phase across all images in our dataset.

\begin{table}[t]
\centering
\footnotesize
\caption{Time required by \fgp and its components.}
\label{tab:rq3-results}
\begin{tabular}{l | c c | c }
\toprule
\textbf{Iteration} &
\makecell{\textbf{Neurons Identifier} \\ \scriptsize{(s)}} &
\makecell{\textbf{Pruner} \\ \scriptsize{(s)}} &
\makecell{\textbf{Total} \\ \scriptsize{(s)}} \\
\midrule
0  & 778.25 & 1.81 & 780.07 \\
10 & 383.04 & 0.92 & 383.96 \\
20 & 358.11 & 0.91 & 359.01 \\
30 & 351.61 & 0.86 & 352.47 \\
40 & 346.72 & 1.01 & 347.73 \\
50 & 345.43 & 0.86 & 346.29 \\
60 & 343.96 & 0.89 & 344.85 \\
70 & 339.30 & 0.91 & 340.22 \\
\midrule
All & 25,777.84 & 64.79 & 25,842.63 \\
\bottomrule
\end{tabular}
\end{table}

\emph{Results}.
\Cref{tab:rq3-results} reports our results. Each row shows the time required by the corresponding iteration, while the last row shows the cumulative runtime across all 70 iterations. 
The \emph{Pruner} (\phase{2}) has a negligible impact on overall execution time, accounting for 0.25\%.
Conversely, \emph{Neurons Identifier} (\phase{1}) accounts for 99.75\% of the execution time.
For performing one iteration, \fgp requires approximately 13 minutes.
Therefore, running 70 iterations requires approximately seven hours. 
This time is reasonable for practical applications, since the pruning is performed offline before deploying the pruned DNN.
Since our general approach allows the use of alternative components, different \emph{Neuron Identifier} (\phase{1}) implementations can be selected if higher efficiency is required.

\begin{Answer}[RQ3 --- Efficiency of \fgp]
The efficiency of \fgp is acceptable for practical applications, given that pruning is performed offline. 
With our dataset, \fgp requires from 13 minutes to a few hours, depending on the selected number of iterations. 
Most of the time is spent extracting the neurons relevant to specific features.
\end{Answer} 

\begin{table*}[t]
\centering
\footnotesize
\caption{Size and effectiveness of the pruned DNN generated by \fgp including misclassified samples.}
\label{tab:rq4-results-misclassified}
\begin{tabular}{l | c c c c | c c c c}
\toprule
& \multicolumn{4}{c|}{\textbf{Size}} & \multicolumn{4}{c}{\textbf{Effectiveness}} \\
\textbf{Iteration} & \makecell{\textbf{FC1}} & \makecell{\textbf{FC2}} & \makecell{\textbf{Params}} & \makecell{\textbf{Size}} & \makecell{\textbf{Accuracy}} & \makecell{\textbf{Precision}} & \makecell{\textbf{Recall}} & \makecell{\textbf{F1-score}} \\
& \scriptsize{(neurons)} & \scriptsize{(neurons)} & \scriptsize{(M)} & \scriptsize{(MB)} & \scriptsize{(\%)} & \scriptsize{(\%)} & \scriptsize{(\%)} & \scriptsize{(\%)} \\
\midrule
\vgg & 4096 & 4096 & 143.67 & 574.70 & 84.79 & 90.74 & 77.10 & 83.22 \\
1 & 1038 & 766 & 47.63 & 190.55 & 70.81 & 90.73 & 64.28 & 74.42 \\
2 & 798 & 562 & 41.06 & 164.26 & 61.84 & 90.75 & 56.04 & 67.10 \\
3 & 720 & 463 & 38.89 & 155.57 & 56.75 & 90.61 & 51.40 & 62.63 \\
4 & 693 & 412 & 38.11 & 152.47 & 54.56 & 90.72 & 49.49 & 61.15 \\
10 & 664 & 298 & 37.18 & 148.75 & 42.43 & 90.66 & 38.49 & 49.96 \\
15 & 664 & 276 & 37.14 & 148.60 & 40.48 & 90.73 & 36.74 & 47.82 \\
16 & 664 & 274 & 37.14 & 148.59 & 41.64 & 90.72 & 37.78 & 48.90 \\
17 & 664 & 271 & 37.14 & 148.57 & 41.37 & 90.64 & 37.53 & 48.37 \\
18 & 664 & 270 & 37.13 & 148.56 & 41.30 & 90.64 & 37.46 & 48.29 \\
19 & 664 & 269 & 37.13 & 148.56 & 40.84 & 90.61 & 37.03 & 47.72 \\
20 & 664 & 268 & 37.13 & 148.55 & 40.45 & 90.61 & 36.67 & 47.51 \\
\bottomrule
\end{tabular}
\end{table*}

\begin{figure*}
    \centering
    \subfloat[Size of pruned DNN.]{
        \begin{tikzpicture}
\begin{axis}[
    width=0.48\textwidth,
    height=0.3\textwidth,
    /pgf/number format/1000 sep={'},
    xlabel={Iteration},
    ylabel={Params (M)},
    ymin=0, ymax=150,
    xmin=0, xmax=20,
    grid=major,
    mark size=1pt,
    legend pos=north east,
    legend cell align={left},
]

\addplot[
        mark=*, blue
    ] coordinates {
    (0,143.67) (1,94.35) (2,78.85) (3,71.52) (4,66.38) (5,62.7) (6,60.12) (7,58.05) (8,56.18) (9,55.02) (10,53.6) (11,52.5) (12,51.61) (13,50.99) (14,50.32) (15,49.74) (16,49.25) (17,48.74) (18,48.16) (19,47.68) (20,47.1) 
};
\addlegendentry{\fgp}

\addplot[
        mark=*, red
    ] coordinates {
    (0,143.67) (1,47.63) (2,41.06) (3,38.89) (4,38.11) (5,37.72) (6,37.53) (7,37.4) (8,37.28) (9,37.2) (10,37.18) (11,37.17) (12,37.16) (13,37.16) (14,37.15) (15,37.14) (16,37.14) (17,37.14) (18,37.13) (19,37.13) (20,37.13)
};
\addlegendentry{With Misclassified}

\end{axis}
\end{tikzpicture}         \label{fig:rq4-size-iterations}
    }
    \hfill
    \subfloat[Accuracy of pruned DNN.]{
        \begin{tikzpicture}
\begin{axis}[
    width=0.48\textwidth,
    height=0.3\textwidth,
    /pgf/number format/1000 sep={'},
    xlabel={Iteration},
    ylabel={Accuracy (\%)},
    ymin=35, ymax=90,
    xmin=0, xmax=20,
    grid=major,
    mark size=1pt,
    legend pos=north west,
    legend cell align={left},
]

\addplot[
        mark=*, blue
    ] coordinates {
    (0,84.7900113507377) (1,81.7063942489595) (2,80.6469920544835) (3,79.2281498297389) (4,78.92546348846) (5,78.5281876655315) (6,77.4120317820658) (7,77.5255391600454) (8,77.8660612939841) (9,77.7147181233446) (10,78.3390087022323) (11,78.4335981838819) (12,78.7930382141505) (13,78.1309118426031) (14,77.6958002270147) (15,77.3741959894059) (16,77.3363601967461) (17,77.4120317820658) (18,77.3174423004162) (19,77.1660991297767) (20,76.4850548618993) 
};
\addlegendentry{\fgp}

\addplot[
        mark=*, red
    ] coordinates {
    (0,84.7900113507377) (1,70.8096859629209) (2,61.842603102535) (3,56.7536889897843) (4,54.5592130155126) (5,52.8187665531592) (6,51.7972001513431) (7,49.8675747256905) (8,45.8948164964056) (9,44.0408626560726) (10,42.4328414680287) (11,42.8679530836171) (12,42.6409383276579) (13,41.9409761634506) (14,41.5247824441922) (15,40.4842981460461) (16,41.6382898221717) (17,41.3734392735527) (18,41.297767688233) (19,40.8437381763147) (20,40.4464623533863)
};
\addlegendentry{With Misclassified}

\end{axis}
\end{tikzpicture}         \label{fig:rq4-accuracy-iterations}
    }
    \caption{Size and effectiveness of the pruned DNN when misclassified inputs are discarded or considered by \fgp.}
    \label{fig:prunedMisclassified}
   \Description{The figure presents two plots, one next to the other.
   In the plot on the left, the x-axis shows the number of iterations (from 0 to 20), and the y-axis shows the number of parameters in millions (from 0 to 150). The plot shows two lines: a blue line for the original \fgp and a red line for the modified version of \fgp with misclassified images.  The plot shows that the original \fgp preserves slightly more parameters than the version with misclassified images for all iterations, with a difference of approx. 10 million parameters in the last iteration.
   In the plot on the right, the x-axis shows the number of iterations (from 0 to 20), and the y-axis shows the accuracy (from 35\% to 90\%). As for the other plot, we are comparing the original \fgp approach (blue) with the alternative version that also uses misclassified inputs (red). The plot shows that the original \fgp significantly outperforms the modified version, with an accuracy difference of almost 40\% at the last iteration.}
\end{figure*}
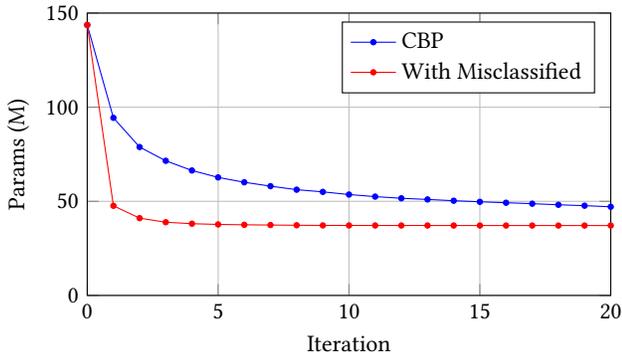
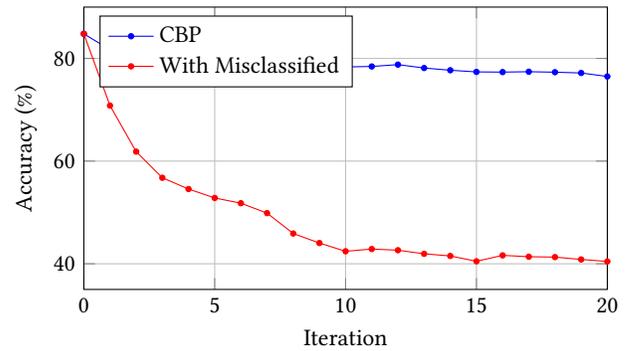

\subsection{Impact of Configuration Options (RQ4)}
\label{sec:rq4}
To assess how different configuration options affect the effectiveness of the resulting pruned DNN, we performed two experiments:
\begin{enumerate}[start=1,label={\bfseries Exp\arabic*:}]
    \item We compare the effectiveness of pruned DNN when samples misclassified by the original DNN are treated as correct or discarded.  
    \item We compare \fga and \efga. For \efga, we also analyzed different aggregation policies. 
\end{enumerate}
\noindent We present the two experiments and their results below. 

\textbf{Exp~1}. We used the same metrics as for RQ1, since our goal is to assess the effectiveness of the pruned DNN.

\emph{Results}. 
\Cref{tab:rq1-results} and \Cref{tab:rq4-results-misclassified} present the effectiveness of the pruned DNN when samples misclassified by the original DNN are discarded or treated as correct.  
\fgp reaches a plateau and stops removing neurons after 70 and 20 iterations, respectively.

Our results show that including misclassified samples triggers a significantly more aggressive pruning strategy compared to the baseline \fgp. 
For example, in the first iteration, considering the misclassified samples enables pruning 74.66\% of neurons from FC1 and 81.30\% from FC2, whereas not considering them enables pruning 35.99\% from FC1 and 42.46\% from FC2. 
For the configuration that considers misclassified samples, this pruning corresponds to a drop in model size from 574.70 MB to 190.55 MB (-66.84\%), while the configuration that does not consider misclassified samples yields a drop in model size from 574.70 MB to 377.42 MB (-34.33\%).
A more aggressive pruning strategy can be beneficial when there are significant resource constraints in which the DNN component must be deployed. 

Furthermore, our results show that including misclassified samples results in a larger performance reduction across iterations than the default implementation.
Specifically, including misclassified samples results in a 13.98\% drop in accuracy during the first iteration. 
This drop is more severe than the standard \fgp experienced even after 70 iterations.
Similar to the default configuration, when configured with misclassified samples, our solution progressively starts pruning fewer neurons. 
By the 20th iteration, the pruned model reached an accuracy of only 40.45\% (-44.34\%) and achieved total reductions of 83.79\% and 93.46\% in the number of neurons in FC1 and FC2, respectively.
Also, the other metrics exhibit the same behavior as in the default configuration: Recall and F1-score decrease, while Precision remains stable.

To visualize our results, \Cref{fig:prunedMisclassified} plots the number of parameters (\Cref{fig:rq4-size-iterations}) and accuracy (\Cref{fig:rq4-accuracy-iterations}) of the pruned DNN when misclassified inputs are discarded (standard implementation) or considered during the pruning.
The results show that including misclassified inputs enables \fgp to reach a plateau more quickly and to remove more neurons.
However, including misclassified inputs results in a significant drop in the pruned~DNN's accuracy.

\begin{figure}
    \centering
    \begin{tikzpicture}
\begin{axis}[
    width=0.39\textwidth,
    height=0.3\textwidth,
    xlabel={Iteration},
    ylabel={Accuracy (\%)},
    /pgf/number format/1000 sep={'},
    ymin=0, ymax=90,
    xmin=0, xmax=25,
    grid=major,
    legend pos=outer north east,
    legend cell align={left},
    legend style={
        font=\scriptsize,
    }
]

\addplot[dash pattern=on 4pt off 1pt, color=blue, mark=none, thick] coordinates {
    (0,84.7900113507377) (1,35.4521377222852) (2,3.63223609534619) (3,0) (4,0) (5,0) (6,0) (7,0) (8,0) (9,0) (10,0) (11,0) (12,0) (13,0) (14,0) (15,0) (16,0) (17,0) (18,0) (19,0) (20,0) (21,0) (22,0) (23,0) (24,0) (25,0)
};
\addlegendentry{TOP(1)}

\addplot[dash pattern=on 4pt off 1pt, color=red, mark=none, thick] coordinates {
    (0,84.7900113507377) (1,54.9186530457813) (2,25.1040484298146) (3,18.936814226258) (4,10.0454029511918) (5,0.58645478622777) (6,0) (7,0) (8,0) (9,0) (10,0) (11,0) (12,0) (13,0) (14,0) (15,0) (16,0) (17,0) (18,0) (19,0) (20,0) (21,0) (22,0) (23,0) (24,0) (25,0)
};
\addlegendentry{TOP(3)}

\addplot[dash pattern=on 4pt off 1pt, color=green!70!black, mark=none, thick] coordinates {
    (0,84.7900113507377) (1,64.3586833144154) (2,44.6651532349602) (3,26.6363980325387) (4,21.1502080968596) (5,11.6723420355656) (6,2.64850548618993) (7,0.8134695421869) (8,0) (9,0) (10,0) (11,0) (12,0) (13,0) (14,0) (15,0) (16,0) (17,0) (18,0) (19,0) (20,0) (21,0) (22,0) (23,0) (24,0) (25,0)
};
\addlegendentry{TOP(5)}

\addplot[dash pattern=on 4pt off 1pt, color=black!50, mark=none, thick] coordinates {
    (0,84.7900113507377) (1,70.1853953840333) (2,63.6019674612183) (3,54.0484298146046) (4,50.0756715853197) (5,43.132803632236) (6,39.1600454029511) (7,35.3575482406356) (8,27.1282633371169) (9,28.3011729095724) (10,26.1066969353007) (11,20.4880817253121) (12,16.0612939841089) (13,12.7317442300416) (14,11.2561483163072) (15,8.75898600075671) (16,7.90768066590995) (17,7.49148694665153) (18,2.23231176693151) (19,0.37835792659856) (20,0) (21,0) (22,0) (23,0) (24,0) (25,0)
};
\addlegendentry{TOP(10)}

\addplot[color=black!50, mark=none, thick] coordinates {
    (0,84.7900113507377) (1,62.5993189557321) (2,58.1157775255391) (3,53.5187287173666) (4,51.1539916761256) (5,49.716231555051) (6,47.2569050321604) (7,43.8138479001135) (8,43.7570942111237) (9,40.6923950056753) (10,41.0707529322739) (11,39.5005675368898) (12,39.0654559213015) (13,39.0087022323117) (14,38.3844116534241) (15,36.2656072644721) (16,36.6250472947408) (17,34.733257661748) (18,34.2413923571698) (19,34.714339765418) (20,32.084752175558) (21,32.1604237608777) (22,31.4226258040105) (23,31.7631479379493) (24,31.3469542186908) (25,30.3821415058645)
};
\addlegendentry{REC(80)}

\addplot[color=blue, mark=none, thick] coordinates {
    (0,84.7900113507377) (1,69.977298524404) (2,66.4964055996973) (3,63.2236095346197) (4,62.1831252364737) (5,60.3291713961407) (6,60.2156640181611) (7,58.5887249337873) (8,58.2292849035187) (9,57.623912220961) (10,56.1483163072266) (11,54.8997351494513) (12,54.1241013999243) (13,54.4835414301929) (14,52.6674233825198) (15,52.4404086265607) (16,51.5891032917139) (17,50.0378357926598) (18,49.451381006432) (19,49.451381006432) (20,48.8460083238743) (21,49.0730230798335) (22,48.6189935679152) (23,47.7298524404086) (24,47.0298902762012) (25,44.6462353386303)
};
\addlegendentry{REC(85)}

\addplot[color=red, mark=none, thick] coordinates {
    (0,84.7900113507377) (1,76.7309875141884) (2,73.0987514188422) (3,70.6772606886114) (4,68.6908815739689) (5,68.5017026106696) (6,67.2909572455543) (7,65.9667045024593) (8,66.042376087779) (9,65.5315928868709) (10,64.3965191070752) (11,64.3965191070752) (12,63.9992432841468) (13,64.3019296254256) (14,64.737041241014) (15,64.737041241014) (16,65.6451002648505) (17,64.7559591373439) (18,65.3045781309118) (19,65.6451002648505) (20,65.6072644721907) (21,65.6640181611804) (22,65.4937570942111) (23,65.8153613318199) (24,65.4559213015512) (25,65.5694286795308)
};
\addlegendentry{REC(90)}

\addplot[color=green!70!black, mark=none, thick] coordinates {
    (0,84.7900113507377) (1,80.7037457434733) (2,78.1309118426031) (3,79.1713961407491) (4,77.9038970866439) (5,77.620128641695) (6,77.8660612939841) (7,77.6390465380249) (8,77.9417328793038) (9,77.9038970866439) (10,78.1119939462731) (11,78.1876655315928) (12,77.7147181233446) (13,77.8471433976541) (14,77.884979190314) (15,77.5066212637154) (16,77.0147559591373) (17,76.4093832765796) (18,76.5228906545592) (19,75.5013242527431) (20,75.766174801362) (21,75.3688989784336) (22,75.4824063564131) (23,75.0851305334846) (24,74.9337873628452) (25,74.9337873628452)
};
\addlegendentry{REC(95)}

\addplot[color=black, mark=none, thick] coordinates {
    (0,84.7900113507377) (1,68.8989784335981) (2,60.1399924328414) (3,55.1835035944003) (4,54.6159667045024) (5,48.675747256905) (6,47.2190692395005) (7,43.0192962542565) (8,40.9004918653045) (9,33.2576617480136) (10,32.4441922058267) (11,27.8471433976541) (12,22.7203934922436) (13,16.6288308740068) (14,16.1937192584184) (15,14.0559969731365) (16,7.56715853197124) (17,6.84827847143397) (18,4.46462353386303) (19,2.81876655315928) (20,0.73779795686719) (21,0.77563374952705) (22,0) (23,0) (24,0) (25,0)
};
\addlegendentry{AVG}

\addplot[dash pattern=on 1pt off 1pt, color=black, mark=none, thick] coordinates {
    (0,84.7900113507377) (1,81.7063942489595) (2,80.6469920544835) (3,79.2281498297389) (4,78.92546348846) (5,78.5281876655315) (6,77.4120317820658) (7,77.5255391600454) (8,77.8660612939841) (9,77.7147181233446) (10,78.3390087022323) (11,78.4335981838819) (12,78.7930382141505) (13,78.1309118426031) (14,77.6958002270147) (15,77.3741959894059) (16,77.3363601967461) (17,77.4120317820658) (18,77.3174423004162) (19,77.1660991297767) (20,76.4850548618993) (21,76.5039727582292) (22,76.5039727582292) (23,76.579644343549) (24,76.3904653802497) (25,76.3715474839197)
};
\addlegendentry{Default}

\end{axis}
\end{tikzpicture}     \caption{Accuracy of different \efga aggregation policies.}
    \label{fig:rq5-accuracy-iterations}
    \Description{The plot shows the performances of different \efga aggregation policies.
    The x-axis shows the number of iterations (0-25), and the y-axis shows the accuracy (0-90\%).
    The plot shows ten different lines: nine for the different \efga aggregation policies (TOP(1), TOP(3), TOP(5), TOP(10), REC(80), REC(85), REC(90), REC(95), and AVG) and one for the original \fgp approach.
    For most aggregation approaches, the accuracy degrades with only a few iterations, reaching 0\%: TOP(1), TOP(3), TOP(5), TOP(10), and AVG.
    For REC(80), REC(85), and REC(90), the accuracy never reaches 0\%, but they perform significantly worse than the original approach.
    Only REC(95) shows a similar behaviour to \fgp, and it even outperforms it slightly in some iterations.}
\end{figure}
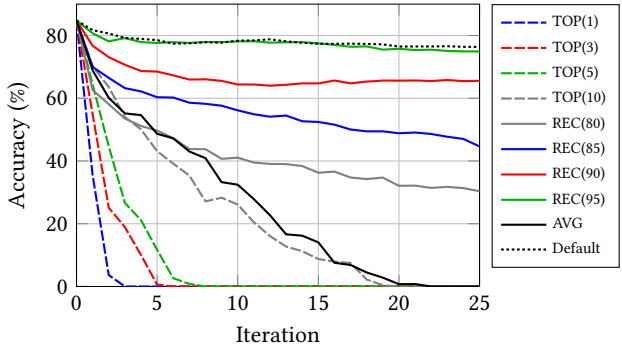

\textbf{Exp~2}. To assess how the use of \efga and its aggregation policies affect the effectiveness of the pruned network, we proceeded as follows.
We consider \efga and different aggregation policies: TOP(1), TOP(3), TOP(5), TOP(10), REC(80), REC(85), REC(90), REC(95), and AVG.
We used the Accuracy of the pruned network to select the best \efga configuration.
Then, we compare this configuration with our default configuration (\fga).
Given the results of Exp~1, we did not include misclassified samples in this experiment.

\begin{table*}[t]
\centering
\footnotesize
\caption{Size and effectiveness of the pruned DNN generated by \fgp with the criterion REC(95).}
\label{tab:rq4-results-rec95}
\begin{tabular}{l | c c c c | c c c c}
\toprule
& \multicolumn{4}{c|}{\textbf{Size}} & \multicolumn{4}{c}{\textbf{Effectiveness}} \\
\textbf{Iteration} & \makecell{\textbf{FC1}} & \makecell{\textbf{FC2}} & \makecell{\textbf{Params}} & \makecell{\textbf{Size}} & \makecell{\textbf{Accuracy}} & \makecell{\textbf{Precision}} & \makecell{\textbf{Recall}} & \makecell{\textbf{F1-score}} \\
& \scriptsize{(neurons)} & \scriptsize{(neurons)} & \scriptsize{(M)} & \scriptsize{(MB)} & \scriptsize{(\%)} & \scriptsize{(\%)} & \scriptsize{(\%)} & \scriptsize{(\%)} \\
\midrule
\vgg & 4096 & 4096 & 143.67 & 574.70 & 84.79 & 90.74 & 77.10 & 83.22 \\
1 & 1349 & 1070 & 56.38 & 225.57 & 80.70 & 90.38 & 73.43 & 80.64 \\
5 & 789 & 628 & 40.94 & 163.81 & 77.62 & 90.11 & 70.65 & 78.38 \\
10 & 636 & 527 & 36.84 & 147.41 & 78.11 & 89.87 & 71.10 & 78.45 \\
15 & 564 & 467 & 34.91 & 139.65 & 77.51 & 89.93 & 70.55 & 77.98 \\
20 & 528 & 444 & 33.95 & 135.83 & 75.77 & 89.71 & 68.97 & 76.86 \\
25 & 510 & 422 & 33.46 & 133.86 & 74.93 & 89.63 & 68.21 & 76.36 \\
\bottomrule
\end{tabular}
\end{table*}

\begin{figure*}
    \centering
    \subfloat[Size of pruned DNN.]{
        \begin{tikzpicture}
\begin{axis}[
    width=0.48\textwidth,
    height=0.3\textwidth,
    /pgf/number format/1000 sep={'},
    xlabel={Iteration},
    ylabel={Params (M)},
    ymin=0, ymax=150,
    xmin=0, xmax=25,
    grid=major,
    mark size=1pt,
    legend pos=north east,
    legend cell align={left},
]

\addplot[
        mark=*, blue
    ] coordinates {
    (0,143.67) (1,94.35) (2,78.85) (3,71.52) (4,66.38) (5,62.7) (6,60.12) (7,58.05) (8,56.18) (9,55.02) (10,53.6) (11,52.5) (12,51.61) (13,50.99) (14,50.32) (15,49.74) (16,49.25) (17,48.74) (18,48.16) (19,47.68) (20,47.1) (21,46.71) (22,46.35) (23,46.06) (24,45.76) (25,45.45)
};
\addlegendentry{Default}

\addplot[
        mark=*, red
    ] coordinates {
    (0,143.67) (1,56.38) (2,47.76) (3,44.42) (4,42.48) (5,40.94) (6,39.65) (7,38.88) (8,38.3) (9,37.54) (10,36.84) (11,36.46) (12,36.08) (13,35.55) (14,35.26) (15,34.91) (16,34.54) (17,34.36) (18,34.24) (19,34.06) (20,33.95) (21,33.87) (22,33.76) (23,33.65) (24,33.59) (25,33.46)
};
\addlegendentry{REC(95)}

\end{axis}
\end{tikzpicture}         \label{fig:rq5-size-iterations}
    }
    \hfill
    \subfloat[Accuracy of pruned DNN.]{
        \begin{tikzpicture}
\begin{axis}[
    width=0.48\textwidth,
    height=0.3\textwidth,
    /pgf/number format/1000 sep={'},
    xlabel={Iteration},
    ylabel={Accuracy (\%)},
    ymin=70, ymax=90,
    xmin=0, xmax=25,
    grid=major,
    mark size=1pt,
    legend pos=north east,
    legend cell align={left},
]

\addplot[
        mark=*, blue
    ] coordinates {
    (0,84.7900113507377) (1,81.7063942489595) (2,80.6469920544835) (3,79.2281498297389) (4,78.92546348846) (5,78.5281876655315) (6,77.4120317820658) (7,77.5255391600454) (8,77.8660612939841) (9,77.7147181233446) (10,78.3390087022323) (11,78.4335981838819) (12,78.7930382141505) (13,78.1309118426031) (14,77.6958002270147) (15,77.3741959894059) (16,77.3363601967461) (17,77.4120317820658) (18,77.3174423004162) (19,77.1660991297767) (20,76.4850548618993) (21,76.5039727582292) (22,76.5039727582292) (23,76.579644343549) (24,76.3904653802497) (25,76.3715474839197)
};
\addlegendentry{Default}

\addplot[
        mark=*, red
    ] coordinates {
    (0,84.7900113507377) (1,80.7037457434733) (2,78.1309118426031) (3,79.1713961407491) (4,77.9038970866439) (5,77.620128641695) (6,77.8660612939841) (7,77.6390465380249) (8,77.9417328793038) (9,77.9038970866439) (10,78.1119939462731) (11,78.1876655315928) (12,77.7147181233446) (13,77.8471433976541) (14,77.884979190314) (15,77.5066212637154) (16,77.0147559591373) (17,76.4093832765796) (18,76.5228906545592) (19,75.5013242527431) (20,75.766174801362) (21,75.3688989784336) (22,75.4824063564131) (23,75.0851305334846) (24,74.9337873628452) (25,74.9337873628452)
};
\addlegendentry{REC(95)}

\end{axis}
\end{tikzpicture}         \label{fig:rq5-accuracy-iterations-rec95}
    }
    \vspace{-.9em}
    \caption{Size and effectiveness of the pruned DNN for the \fgp default configuration and \fgp with \efga and REC(95).}
    \label{fig:fgaANDefgaComparison}
    \Description{The figure presents two plots, one next to the other.
    In the plot on the left, the x-axis shows the number of iterations (from 0 to 25), and the y-axis shows the number of parameters in millions (from 0 to 150). The plot shows two lines: a blue line for the original \fgp and a red line for \efga with REC(95).  The plot shows that the original \fgp preserves slightly more parameters than the REC(95) version at all iterations, with a difference of approx. 10 million parameters in the last iteration.
    In the plot on the right, the x-axis shows the number of iterations (0-25), and the y-axis shows the accuracy (70-90\%). As for the other plot, we are comparing the original \fgp approach (blue) with \efga with REC(95) (red). The plot shows that the behaviour of the two approaches is very similar, with \fgp slightly outperforming REC(95) for most iterations.}
\end{figure*}
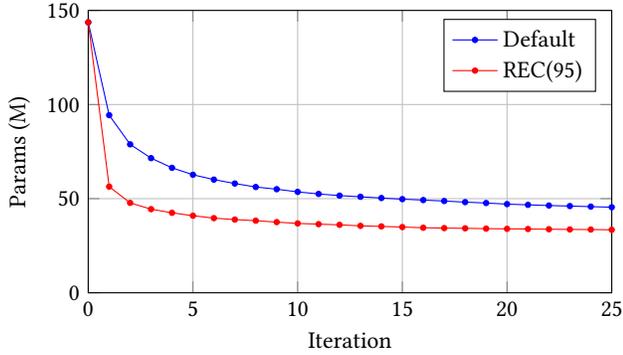
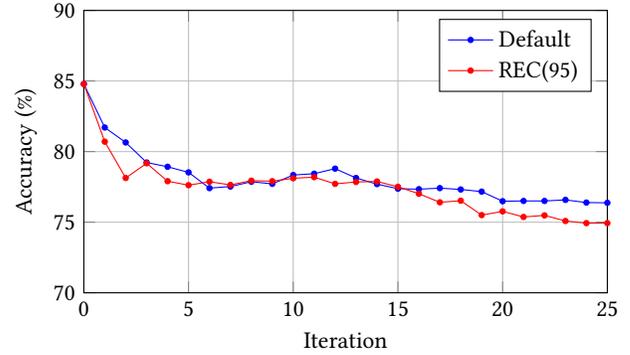

\emph{Results}. \Cref{fig:rq5-accuracy-iterations} shows the accuracy of the pruned DNN for different \efga aggregation policies. 
Several criteria, specifically TOP(1), TOP(3), TOP(5), TOP(10), and AVG, proved excessively aggressive. These methods pruned critical neurons too early, leading to rapid collapse in model performance within just a few iterations. The criteria REC(80), REC(85), and REC(90) exhibited a less aggressive pruning strategy, yet still experienced a rapid drop in accuracy.
Considering these results across the different aggregation policies, we selected REC(95) because it offers accuracy comparable to \fgp.
We used this aggregation policy for our comparison.

\Cref{tab:rq1-results} and \Cref{tab:rq4-results-rec95} present the effectiveness of the pruned DNN by our implementation when its default configuration and \efga with REC(95) as aggregation policy are set.
\Cref{fig:fgaANDefgaComparison} plots the parameters (\Cref{fig:rq5-size-iterations}) and accuracy (\Cref{fig:rq5-accuracy-iterations-rec95}) of the pruned DNN for its default configuration and the chosen configuration for \efga.
At the first iteration, \fgp maintains a slight accuracy advantage (+1.00\%) while pruning 2560 fewer neurons than \efga.
Subsequently, \efga generally demonstrates superior effectiveness: for the same volume of pruned neurons, it achieves higher accuracy in fewer iterations than \fgp. For example, when \fgp stopped pruning, it reached an accuracy of 75.24\% with 6772 pruned units, whereas \efga achieved an accuracy of 77.62\% (+2.38\%) with a comparable number (6775) of pruned neurons. 
This suggests that \efga with REC(95) is not only more accurate but also converges faster, making it a more practical choice.
Additionally, neuron selection based on recall-oriented metrics can enable more effective pruning.
In contrast, more aggressive filtering criteria result in immediate degradation of DNN performance, making them unsuitable for fine-grained pruning~tasks.

\begin{Answer}[RQ4 --- Impact of Configuration Options ]
Including misclassified samples enables \fgp to remove more neurons, but the accuracy of the pruned DNN significantly drops.
\efga with REC(95) outperforms the default configuration in producing smaller and more accurate pruned DNNs.
\end{Answer}

\section{Discussion and Threats to Validity}
\label{sec:discussion}
In this section, we discuss the practical implications of our results and present threats to validity.

The results of RQ1 show that while accuracy, recall, and F1-score drop significantly, the precision of the pruned DNN does not decrease. 
This makes the pruning solution particularly suitable for practical applications that require confidence in the presence of a feature; i.e., when the pruned DNN detects a feature, the input shows that feature.
Furthermore, although analyzing the effectiveness of \fgp across different layers was not part of our research questions, the results from RQ1 show that pruning more in the first fully connected layer (FC1) results in \fgp removing more parameters than pruning the second fully connected layer (FC2). 
This result is consistent with the network structure: each FC1 neuron is connected to a high-dimensional flattened feature vector as well as to all neurons in FC2 (25,088 + 4,096 parameters), whereas each FC2 neuron is connected only to FC1 and has a smaller output layer (4,096 + 1,000 parameters). 
Therefore, pruning a neuron from FC1 removes more parameters than pruning a neuron from FC2, since pruning a neuron removes all its input and output connections.
This result suggests that, in practical applications, engineers need to carefully consider which layers to prune.

The results from RQ2 show that, despite the number of neurons within the FC1 and FC2 layers decreasing significantly, the time savings are limited. 
We noticed that this result is reasonable since most of the computational complexity (99.39\%) is from the convolutional layers, and only a small part comes from FC1 (0.52\%) and FC2 (0.09\%).
Unfortunately, our implementation does not enable our solutions to consider other layers of the DNN since \fga is only applicable to feed-forward layers.
We plan to extend our solution to support the pruning of other layer types.

The results from RQ3 show that although the time required by \fgp is reasonable for practical applications, the \emph{Neurons Identifier} (\phase{1}) component requires the highest computational time.
However, in practice, this cost can be mitigated by running the \emph{Neurons Identifier} with dedicated hardware support (e.g., GPUs) or by considering alternative solutions for identifying the relevant neurons.

The analysis of alternative configuration options (RQ4) showed that, when considering misclassified samples, as in the original \fga~\cite{Gopinath_2023} implementation, pruning becomes more aggressive. 
This result is surprising and contrary to our expectations: We expected those samples to activate more neurons (not relevant to correct predictions) and make the decision tree more complex and less accurate.
Instead, the \emph{Neurons Identifier} (\phase{1}) returns a smaller set of neurons since \fga returns fewer rules. 
This makes pruning more aggressive but negatively affects the network's accuracy.

\textbf{Threats to Validity}.
Though our dataset (\rival) and NN architecture (\vgg) are widely known benchmarks, our concept-based pruning solution can provide different results on other datasets or architectures.
The fact that \vgg is large and widely used mitigates this threat.

The choice of the DNN layer to analyze and the configuration of our solution can threaten the \emph{internal validity} of our results.  Considering different layers and configuration options can yield different results. 
To mitigate this threat, we considered two different layers and compared the effectiveness of alternative configurations.  \section{Related Work}
\label{sec:related}
Our related work considers approaches that explain and prune the internal behavior of DNN.

\emph{Explaining}. Concept-based explanation methods aim to explain a DNN's behavior by focusing on a single concept.
A concept is an abstraction, such as a color, an object, or even an idea~\cite{molnar2025}.
Recent surveys~\cite{Lee2025,Poeta2025} classify concept-based explanation methods as follows.
Symbolic concept-based explanation methods are driven by human-defined symbols, such as high-level attributes or interpretable abstractions (e.g., color or shape). 
They require auxiliary data with concept annotations.
Contrarily, unsupervised techniques cluster samples that the network learns autonomously. 
Although they are not built to resemble human-defined concepts, they may still capture human-understandable abstractions and are extracted via clustering algorithms either post-hoc or during training.

A significant body of research investigates the internal dynamics of DNNs by analyzing the functional roles of individual neurons. 
Recent surveys~\cite{Poeta2025,Lee2025} classify these approaches as Post-hoc Concept-based Explanation Methods. 
Such techniques typically aim to identify sparse subsets of neurons that collectively contribute to a specific model prediction or represent high-level semantic features. 
For example, Kim et al.~\cite{Kim2018} introduce the notion of Concept Activation Vectors (CAVs). The core idea is to link the neural network's internal activation space with a space of human-interpretable concepts, enabling interpretation of learned features.
Gopinath et al.~\cite{Prophecy_2019, Prophecy_2025} introduced Prophecy, a property inference technique that derives formal assertions about neuron activation status (``on''/``off'') and extracts rules for correctly vs. misclassified inputs, establishing a foundation for rule-based analysis of DNNs. Building on this, \fga~\cite{Gopinath_2023} extended the framework to accept neuron numerical activations, and subsequently 
Formica et al.~\cite{FGA_Replication} confirmed its robustness by independent replication.

\emph{Pruning}. Concept-Based pruning is a model compression technique. 
Model compression reduces the size of the AI model, thereby lowering computational demand and complexity, while increasing deployability and inference speed without significantly sacrificing predictive accuracy~\cite{Li2023,Dantas2024}. 
Model compression refers to several different approaches, e.g., knowledge distillation, parameter quantization, and model pruning. The latter removes components of a network to minimize the number of parameters without significantly affecting model performance.
A recent taxonomy~\cite{Cheng2024} classified model pruning techniques considering three aspects.
First, it considers whether the technique is structured or unstructured. 
Unstructured pruning zeros out individual weights (e.g., \cite{Frantar2023}), whereas structured pruning removes entire neurons, filters, or channels, along with all their associated weights (e.g., \cite{li2017pruningfiltersefficientconvnets,Ma2023}).
Second, it considers whether the pruning process is applied before (e.g., \cite{wang2020pickingwinningticketstraining,Tanaka2020}), during (e.g., \cite{Evci2020,Huang2018}), or after training (e.g., \cite{Ma2023,Frantar2023}), or at runtime (e.g., \cite{Rao2019,tang2021manifold}).
Finally, it considers whether the pruning is Magnitude-Based, $l_p \text{ Norm}$, Sensitivity (a.k.a. Saliency), and Loss Change.
Magnitude-Based pruning removes parameters with the smallest absolute value, assuming they contribute less to the model output (e.g., \cite{Han2016}).
$l_p \text{ Norm}$-based pruning evaluates the importance of groups of parameters (e.g., \cite{li2017pruningfiltersefficientconvnets,sun2024simpleeffectivepruningapproach}).
Sensitivity (a.k.a. Saliency) studies how sensitive the model performance is to the removal of a parameter and/or how the loss changes when specific weights are pruned (e.g., \cite{Lee2019,Zhao2019}).
Loss Change assesses a parameter's significance by comparing the model's loss with and without it. This is typically done using a Taylor expansion-based approximation (e.g., \cite{Ma2023,Fang2023}).
Considering these aspects, our pruning technique can be classified as (i) both structured and unstructured (depending on the implementation), (ii) after training, and (iii) based on a novel concept-based pruning criterion.

Some solutions explore pruning using the notion of circuits.
A circuit is a sub-network of a larger neural network responsible for one or more specific features~\cite{Olah2020}, and it provides a lens for understanding model behavior. Recent works use pruning to isolate these circuits, producing highly specialized models.
Hamblin et al.~\cite{hamblin2022pruning} propose a saliency-based approach to extract circuits responsible for specific visual features in CNNs. While our goal is resource efficiency and predictive performance, they focus on extracting interpretable circuits from the model.
Anani et al.~\cite{anani2026} presented Certified Circuits, introducing formal guarantees on circuit stability by wrapping any black-box discovery algorithm with randomized data sub-sampling to certify that extracted sub-networks remain consistent under input perturbations and model variations.
Their work produces a circuit for each class, whereas our approach produces a single pruned model with a multiclass output.
Input perturbations are used to generate out-of-distribution samples, which could be used to assess the robustness of DNNs~\cite{Arcaini2020,Arcaini2022,Damiano2025,Amini2020}. 
Bhaskar et al.~\cite{NEURIPS2024_20fdaf67} propose Edge Pruning, a scalable optimization method using gradients to discover circuits. While we use explainability tools to produce smaller DNNs, they use pruning to find circuits and make language models more interpretable.
Unlike these works, our pruning solution is concept-based, using user-chosen, human-understandable features.

To the best of our knowledge, this paper presents the first concept-based pruning solution. Unlike classical numerical solutions, our concept-based pruning framework reduces the size of DNNs by considering semantic concept criteria and is particularly well-suited for effectively integrating large DNNs into systems. \section{Conclusion}
\label{sec:conclusion}
We presented a concept-based pruning framework for DNNs to support their integration by accounting for the target system's requirements, including relevant concepts (e.g., class and feature labels) and performance.
We implemented this framework by reusing existing components (\fga, \efga, and Torch-Pruning).
We evaluated the effectiveness of our solution on the \vgg architecture.
Our empirical results show that concept-based pruning can significantly reduce the number of neurons and parameters, producing smaller, more efficient models.
Although recall decreases in some configurations, precision remains stable, and our solution enables engineers to select different configurations to achieve trade-offs among accuracy, pruning aggressiveness, and computational cost.

\section*{Data Availability}
A complete replication package is available at \cite{replicationFGP}.

\bibliographystyle{ACM-Reference-Format}

\end{document}